\documentclass[aps,pre,reprint, amsmath, amssymb,superscriptaddress]{revtex4-1}
\usepackage{graphicx} 
\usepackage{braket}

\pdfoutput=1
\usepackage[utf8]{inputenc}
\usepackage{bm}
\usepackage[english]{babel}
\usepackage[T1]{fontenc}
\usepackage{mathrsfs}
\usepackage[retainorgcmds]{IEEEtrantools}
\usepackage{amsmath}
\usepackage{amssymb}
\usepackage{color}
\usepackage{amsfonts}
\usepackage{times,txfonts}
\usepackage{nicefrac}
\usepackage[colorlinks=true,linkcolor=blue,urlcolor=blue,citecolor=blue,pdfusetitle]{hyperref}
\usepackage{mathtools}
\usepackage{isotope}
\usepackage{siunitx}
\usepackage{dsfont}
\usepackage[percent]{overpic}

\usepackage{float}
\graphicspath{{images/}}

\usepackage{braket}

\usepackage{lipsum}
\usepackage{enumerate}

\usepackage[dvipsnames]{xcolor}
\usepackage{accents}

\begin{document}
\setlength{\abovedisplayskip}{4pt}
\setlength{\belowdisplayskip}{4pt}

\title{Demonstration of energy extraction gain from non-classical correlations}

\author{A.~Stahl}
	\affiliation{Institut f\"ur Physik, Universit\"at Mainz, Staudingerweg 7, 55128
		Mainz, Germany}
	\author{M.~Kewming}
	\affiliation{School of Physics, Trinity College Dublin, College Green, Dublin 2, Ireland}
	\author{J.~Goold}
	\affiliation{School of Physics, Trinity College Dublin, College Green, Dublin 2, Ireland}
\author{J.~Hilder}
	\affiliation{Institut f\"ur Physik, Universit\"at Mainz, Staudingerweg 7, 55128
		Mainz, Germany}
     \author{U.~G.~Poschinger}\email{poschin@uni-mainz.de}
	\affiliation{Institut f\"ur Physik, Universit\"at Mainz, Staudingerweg 7, 55128
		Mainz, Germany}
    \author{F.~Schmidt-Kaler}
	\affiliation{Institut f\"ur Physik, Universit\"at Mainz, Staudingerweg 7, 55128
		Mainz, Germany}
\date{\today}

\begin{abstract}
Within the framework of microscopic thermodynamics,  correlations can play a crucial role for energy extraction. Our work sheds light on this connection by demonstrating that entanglement governs the amount of extractable energy in a controllable setting. We experimentally investigate a fundamental link between information, encoded in tunable non-classical correlations and quantified by quantum state tomography, and its utility as \textit{fuel} for energy extraction. We realize an agent-demon protocol involving two trapped-ion qubits, and show that by implementing an appropriate feedback policy, the demon can optimize the energy extraction process, capitalizing on the correlations between the system's constituents. By quantifying both the concurrence of the two-qubit resource state and the energy extraction gain from applying the feedback policy, we corroborate the connection between information and energy, solidifying the role of non-classical correlations as a resource for thermodynamic processes.
\end{abstract}

\maketitle


Microscopic thermodynamics is a prospering field in modern physics, studying how concepts and notions from classical thermodynamics carry over to the microscopic domain, where fluctuations of thermal or quantum nature play a significant role. By adapting classical principles of thermodynamics to microscopic systems, researchers have gained invaluable insights into the resources that underpin the operation of microscopic engines and other quantum machines \cite{Goold_2016,Kosloff_2013, Vinjanampathy_2016, Deffner_2019, Landi_2021}. 
This has lead to the identification of specific thermodynamic properties of quantum thermal machines \cite{Scully_2011, Brunner_2014, Mitchison_2015,   Brander_2017,  Kilgour_2018, Dann_2020, Karmi_2016}, 
which operate by manipulating the quantum state of a system to extract energy or perform work. Technological advances with the last two decades have enabled experimenters to extract information from a microscopic system and conditionally act back on it, thus bringing the concept of a \textit{demon} to reality \cite{KOSKI2015, Masuyama2018}. Generalizing the famous Maxwell demon, this notion describes an unsupervised entity acting within a physical system, by acquiring information and acting back on it. Studying the impact of such demons on the properties of physical systems is an active field of research \cite{MARUYAMA2009,Kim2011}.

However, despite the growing body of research on the thermodynamic features of quantum systems, a contentious debate surrounds the specific role of correlations, in particular of non-classical correlations, in energy extraction schemes \cite{Hovhannisyan_2013, Dominik_2023}. The identification of an intricate link between thermodynamics and entanglement \cite{Plenio_2001} has sparked intense interest in understanding the role of non-classical correlations as resources for thermodynamic processes \cite{Oppenheim_2002,Bresque2021}. As highlighted in \cite{Francica_2017}, non-classical correlations provide a lower bound on the extractable energy, but classically correlated states can outperform them, producing higher yields. 

Here, we present the experimental demonstration of a minimal protocol for showing how entanglement can act as a fuel for energy extraction from multipartite systems. It is realized by employing qubits encoded in trapped atomic ions. The strengths of this versatile platform - in particular the availability of precise methods for manipulating and reading out qubits, render it to be ideally suited for applications in microscopic and quantum thermodynamics \cite{Ro_nagel_2016, Schulz_2019, Maslennikov_2019, Onishchenko_2022}. The protocol is based on two parties, a \textit{demon} and an \textit{agent}, each having access to one constituent of a two-qubit state, henceforth referred to as \textit{resource}. The demon's objective is to extract as much energy as possible from the resource, by strategically minimizing the probability of the agent's qubit concluding in the excited state. While the demon possesses knowledge about the initial resource state and the capability to implement joint unitary operations on the system, she lacks the ability to measure the state of the agent's qubit. The demon uses a measurement in conjunction with knowledge about correlations as a resource to acquire information regarding the state of the agent. Subsequently, to extract  maximum energy, she utilizes the acquired information to apply feedback \cite{Groisman_qunatum_2005, Francica_2017}. The question addressed is therefore: by how much can the demon increase the maximum amount of extractable energy from her qubit by means of applying a feedback policy?  Thus, the quantity of interest is the \textit{demonic gain} $\delta\mathcal{W}$, which is the difference of the demon's extracted energy for the case of an optimal feedback policy to the case without. Previous theoretical work has showed that for pure resource states, the demonic gain is bounded from below by the entanglement between the agent's and the demon's subsystems, while it is bounded from above by the energy content of the resource \cite{Francica_2017}. In particular, with the entanglement between the agent's and the demon's subsystems specified by the concurrence $\mathcal{C}$, the lower bound is given by \footnote{The additional factor $1/2$ does not appear in the bound given in Ref. \cite{Francica_2017}, as the free Hamiltonian is trivially different ($\hat{H}_0=-\hat{\sigma}_z)$).} 
\begin{equation}
\delta\mathcal{W} \geq \frac{1}{2}\left(1-\sqrt{1-\mathcal{C}^2}\right)
\label{eq:dWlowerBound}
\end{equation}

\begin{figure*}
\centering
\includegraphics[width=\textwidth,trim={1.8cm 6.0cm 3.0cm 2cm},clip]{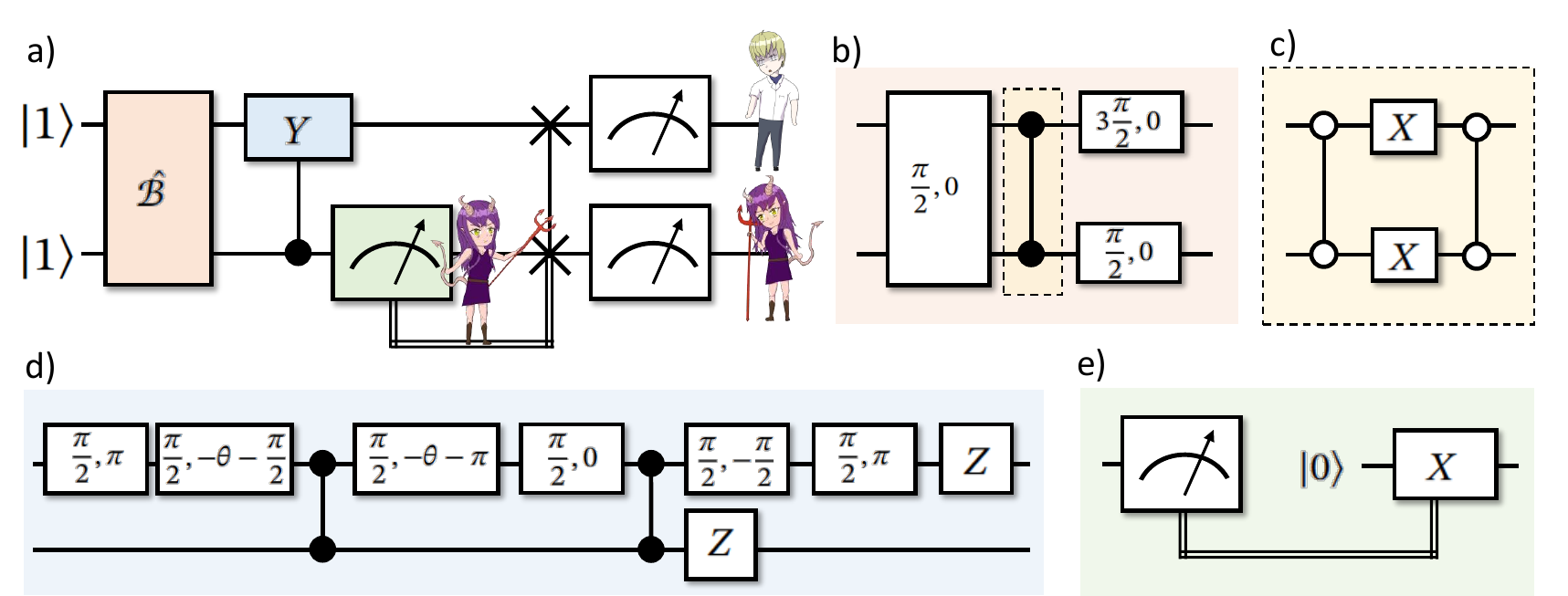}
\caption[]{\textbf{(a)} Quantum circuit for the correlation-enabled work extraction protocol. Both qubits are prepared a Bell state via gate sequence $\hat{\mathcal{B}}$, for which the decomposition into native gates is shown in \textbf{(b)} along with the decomposition of the controlled-$Z$ $\pi/2$ gate in \textbf{(c)}. A parametric controlled-$Y$ gate, with decomposition shown in \textbf{(d)}, is used to prepare the state $\ket{\Psi(\theta)}$. Then, the demon measures her qubit, with the emulation of a QND measurement shown in \textit{(e)}. Depending on the measurement outcome, she performs a controlled-SWAP operation. The two-qubit gate with open circles in c) are controlled-$Z$ $\pi/4$ gates (see text). }
\label{fig:circuits}
\end{figure*}

\textit{Protocol for entanglement-enhanced energy extraction} - The work extraction protocol is described by a quantum circuit operating on two qubits, shown in Fig.~\ref{fig:circuits}. Both qubits are described by a free Hamiltonian $\hat{H}_{0}=\hbar \omega \ket{1}\bra{1}$, where $\ket{1}$ corresponds to the energetically higher state and $\hbar\omega=\epsilon$ is the qubit energy splitting. The initial resource state provided to demon and agent is
\begin{equation}
\ket{\Psi(\theta)} = \frac{1}{\sqrt{2}}\left(\cos (\theta) \ket{1_A0_D}+\ket{0_A1_D} - \sin (\theta) \ket{0_A0_D}\right)
\label{eq:initialstate}
\end{equation}
This state is obtained by preparing a maximally entangled Bell state $\ket{\Psi^+} = (\ket{0_A1_D}+\ket{1_A0_D})/\sqrt{2}$ and applying a parametric controlled-$Y$ gate $ c\hat{Y}(\theta)  = \mathds{1}_A\otimes \ket{1_D}\bra{1_D} + e^{-i \frac{\theta}{2}  \hat{Y}_A } \otimes \ket{0_D}\bra{0_D}$. The angle $\theta$ controls the amount of correlations: for $\theta=0$, the maximally entangled Bell state is retained, while for $\theta=\pi/2$, the agent's and demon's qubits are in a separable state. The entanglement between the agent's and the demon's qubits can be quantified via the concurrence $\mathcal{C}$ \cite{horodecki2009quantum}
\begin{align}
\mathcal{C} = \max \{0, \lambda_{1}-\lambda_{2}-\lambda_{3}-\lambda_{4}\}\,,
\label{eq:concurrence}
\end{align}
where $\lambda_{i}$ represent the eigenvalues of $R=\sqrt{\sqrt{\rho}\tilde{\rho}\sqrt{\rho}}$, with $\tilde{\rho}=(\sigma_{y}\otimes\sigma_{y})\rho^{*}(\sigma_{y}\otimes\sigma_{y})$ and $\rho^{*}$ is the conjugate of $\rho$.
For the initial state Eq. \ref{eq:initialstate}, this yields
\begin{equation}
\label{eq:conc}
\mathcal{C}(\theta) = |\cos {\theta}|
\end{equation}
With a resource state Eq. \ref{eq:initialstate} prepared, the protocol is as follows: The demon performs a projective measurement on her qubit in the energy basis, yielding outcomes $d=0_D$ or $d=1_D$ at balanced probabilities $Pr(d=0_D)=Pr(d=1_D)=1/2$. She uses this information to execute a feedback mechanism, consisting of the application of a SWAP gate between both qubits ($\hat{U}_0$), which is only carried out once measurement result $0_D$ has been obtained, otherwise no operation is carried out ($\hat{U}_1=\mathds{1}$). This policy is optimal, as swapping the qubits is the best course of action for the demon, upon having obtained measurement result $0_D$. With the resource state $\hat{\rho}=\ket{\Psi(\theta)}\bra{\Psi(\theta)}$, the projection operators  $\hat{\Pi}_d=\mathds{1}_A\otimes \ket{d}\bra{d}$ and the post-measurement states
\begin{equation}
\hat{\rho}_{d}=\hat{\Pi}_d\;\hat{\rho}\;\hat{\Pi}_d/\text{Pr}(d),
\end{equation}
the final state after execution of the feedback operation is given by
\begin{eqnarray}
\hat{\rho}' &=& \tfrac{1}{2}\hat{U}_0 \hat{\rho}_{0_D} \hat{U}_0^{\dagger} + \tfrac{1}{2}\hat{U}_1 \hat{\rho}_{1_D} \hat{U}_1^{\dagger}.
\label{eq:finalstate}
\end{eqnarray}
The demonic gain is given by the difference of the initial demon's energy as determined by the resource and the energy extracted after applying the optimal feedback policy: 
\cite{Francica_2017}
\begin{equation}
    \delta \mathcal{W} = \text{Tr}\left[\hat{H}_{0,D} \left(\hat{\rho}'-\hat{\rho}\right)\right]
    \label{eq:deltaW}
\end{equation}
Evaluating Eq. \ref{eq:deltaW} reveals that the demonic gain is given by the concurrence of the resource:
\begin{equation}
\frac{\delta\mathcal{W}}{\epsilon}  = \frac{\mathcal{C}(\theta)^{2}}{2}.
\label{eq:dWvsC}
\end{equation}
This result illuminates a direct connection between the work extracted in the feedback protocol by the demon and the correlations of the resource.

\begin{figure}[htp!]
    \centering
    \includegraphics[width=\columnwidth]{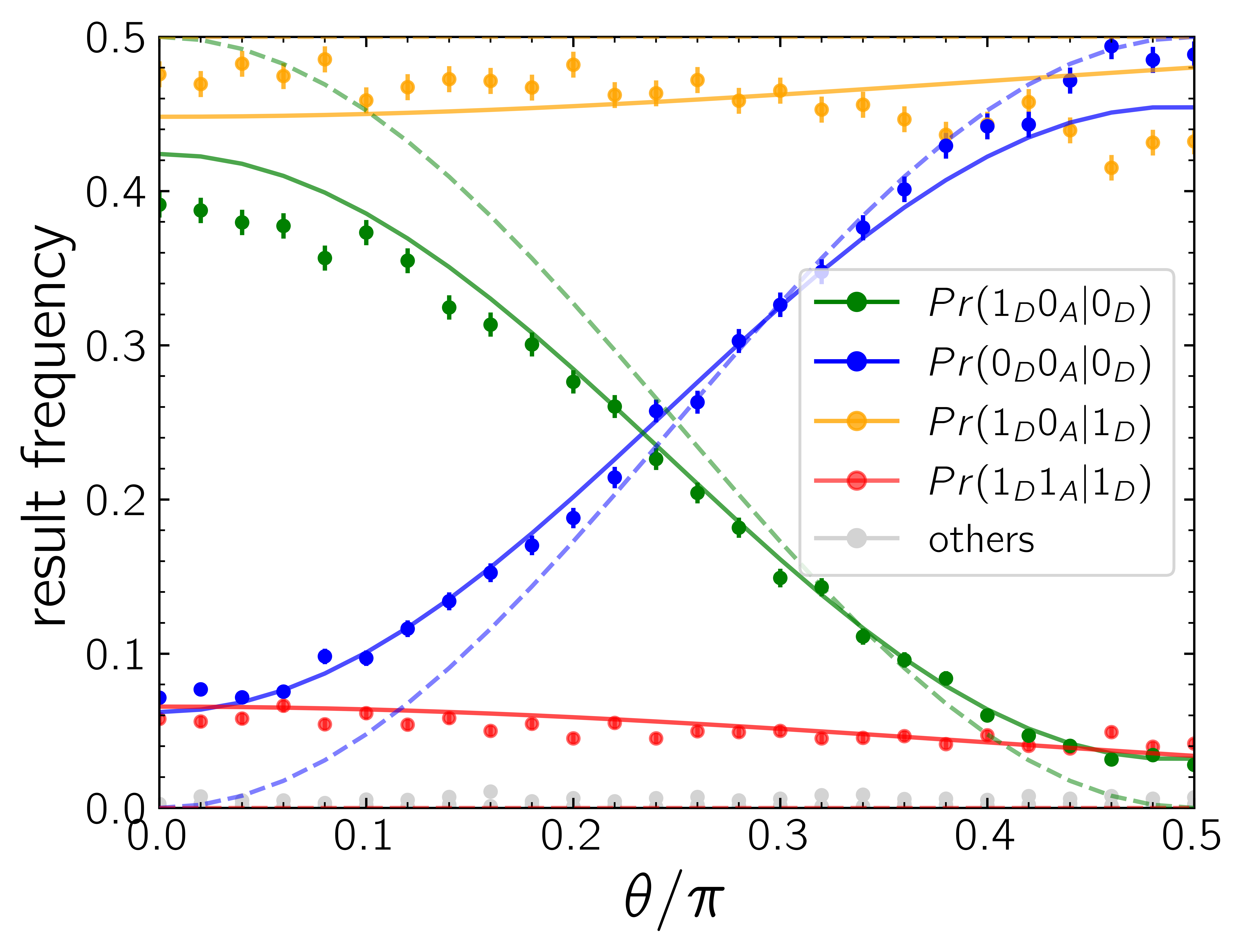}
    \caption{Estimated outcome probabilities $\text{Pr}(d'a'|d)$ versus the preparation angle $\theta$. The  probability that the demon successfully extracts work after swapping her qubit with the agent's one (green) decreases with $\theta$, while the probability of energy extraction by neither party (blue) increases with $\theta$. The probability of initial success by the demon and no swap operation (yellow) remains constant, and spurious work extraction by both parties (red) is due gate errors (see text). Further outcomes (cumulated, grey) are due to preparation and measurement errors. Each data point corresponds to 3500 independent and identical repetitions of the protocol. The ideal results (dashed) and the results expected from modelling imperfect two-qubit gates (solid) are shown, see text for details.
    }
    \label{fig:prob_theta}
\end{figure}

\textit{Experimental realization - } The qubit platform employed for this work is atomic $\isotope[40]{Ca}^{+}$ ions stored in a segmented Paul trap \cite{Schulz_2008}. The qubits are encoded in the spin of the ion's valence electron \cite{POSCHINGER2009, Onishchenko_2022}. The qubit states pertain to the spin's alignment to an external magnetic field, yielding an energy splitting of $\epsilon/\hbar \approx 2\pi\times\SI{10}{\mega\hertz}$. The qubits are initialized via optical pumping and read out via population transfer from state $\ket{0}$ to the metastable $D_{5/2}$ state, while the detection relies on recording state-dependent resonance fluorescence \cite{POSCHINGER2009}. All laser-driven operations take place at a fixed trap site, the laser interaction zone (LIZ). The ions are shuttled within the segmented trap by supplying suitable electric waveforms to the trap electrodes \cite{Kaushal2020}, such that either only the demon's qubit, or the only agent's qubit, or both at the same time can undergo laser-driven operations or qubit readout. Qubit rotations of the form $\hat{R}(\theta,\phi)=\exp\left[-\tfrac{i}{2}\theta \left(\cos\phi\hat{X}+\sin\phi\hat{Y}\right)\right]$, including Pauli gates,  are driven by a pair of co-propagating far off-resonant laser beams generating resonant stimulated Raman transitions. Two-qubit controlled-$Z$ gates of the form
$\hat{U}_{ZZ}=\exp\left[-i\frac{\pi}{8}\hat{Z}\otimes\hat{Z}\right]$
are driven by a pair of counter-propagating far off-resonant laser beams \cite{LEIBFRIED2003A,Hilder2022}. The initial odd-parity Bell state $\ket{\Psi^+}$ is prepared by simultaneously preparing both qubits in superposition states, acting with two $\hat{U}_{ZZ}$ gates interspersed with a simultaneous Pauli-$X$ gates for suppressing gate errors
 \cite{BALLANCE2014}, see Fig. \ref{fig:circuits} b). The preparation of the resource state is concluded by a controlled-$Y$ gate with variable rotation angle, which is decomposed in two $\hat{U}_{ZZ}$ gates and local rotations as shown in Fig. \ref{fig:circuits} d).\\

The protocol requires real-time feedback operations, conditioned on mid-sequence measurements. These  operations are carried out by i) accumulating photon counts on a system-on-a-chip (SoC) device, ii) comparing the count number to a specified threshold on the SoC to discriminate measurement results '0' (count below threshold) and '1' (count above threshold), iii) a real-time logic on the SoC deciding which stored voltage ramps are to be generated and sent to the trap electrodes. \\ As the demon's first qubit measurement via state-dependent fluorescence is completely depolarizing, an ideal QND measurement has to be realized via feedback as shown in Fig. \ref{fig:circuits} e): After the measurement, the demon's qubit is re-initialized to $\ket{0_D}$ via optical pumping. If the qubit has previously detected in $\ket{1_D}$ it is kept in the LIZ, otherwise it is moved away. In any case, laser pulses for a Pauli-$X$ gate are carried out, such that the state is only flipped to $\ket{1_D}$ for previous detection in $\ket{1_D}$. The conditional SWAP gate enacted by the demon upon detecting her qubit in $\ket{0_D}$ is driven by voltage ramps which rotate the two ion Coulomb crystal such that their identities are swapped \cite{KAUFMANN2014}. This process is carried out only once the demon's qubit has been previously measured in $\ket{1_D}$. 

\begin{figure}[htp!]
    \centering
     \includegraphics[width=\columnwidth,trim={9cm 1cm 9cm 1.5cm},clip]{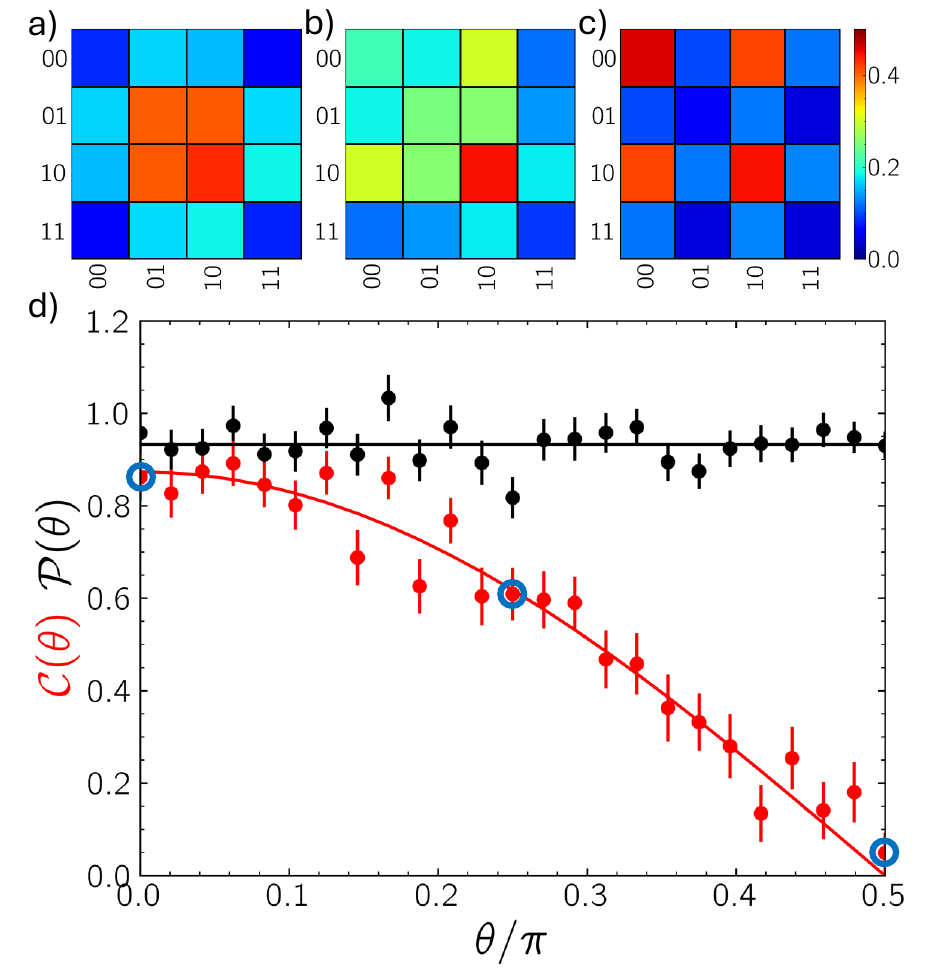}
    \caption{Characterization of the resource state: d) Measured concurrence (red) and purity (black) versus the preparation angle $\theta$. For each data point, quantum state tomography with 100 shots per Pauli operator was carried out, the confidence intervals are computed via parametric bootstrapping (see text). The top panels (a), (b), and (c) show reconstructed density matrices (absolute values) for the values $\theta=\pi \times \{0,0.25,0.5\}$, respectively.}
    \label{fig:fig3}
\end{figure}

\textit{Results -} The protocol is executed with the preparation angle $\theta$ varying between 0 and $\pi/2$. For each value of $\theta$, independent repetitions of the protocol are carried out. We estimate the joint probabilities for each final agent and demon outcomes $\text{Pr}(d',a'|d)$, conditioned on the demons mid-circuit outcome $d$, by the relative frequencies for each outcome. The measured outcome probabilities versus $\theta$ are shown in Fig.~\ref{fig:prob_theta}. It can be seen that the probability $\text{Pr}(1_D0A|0_D)$ of the demon to successfully increase the amount of energy extracted from the resource by swapping the qubits decreases with the preparation angle $\theta$, or respectively with decreasing correlations of the resource. The demon's final extracted energy is obtained by the marginal probability of measurement outcome $d_D'=1_D$:
\begin{equation}
\frac{\mathcal{W}_f}{\epsilon} = \text{Pr}(d_D'=1_D).
\label{eq:Wf}
\end{equation}
Likewise, the energy the demon could extract without applying feedback is determined by the first measurement outcome:
\begin{equation}
\frac{\mathcal{W}_i}{\epsilon} = \text{Pr}(d_D=1_D).
\label{eq:Wi}
\end{equation}

\begin{figure}[htp!]
    \centering
    \includegraphics[width=\columnwidth,trim={3cm 0.5cm 3cm 1cm},clip]{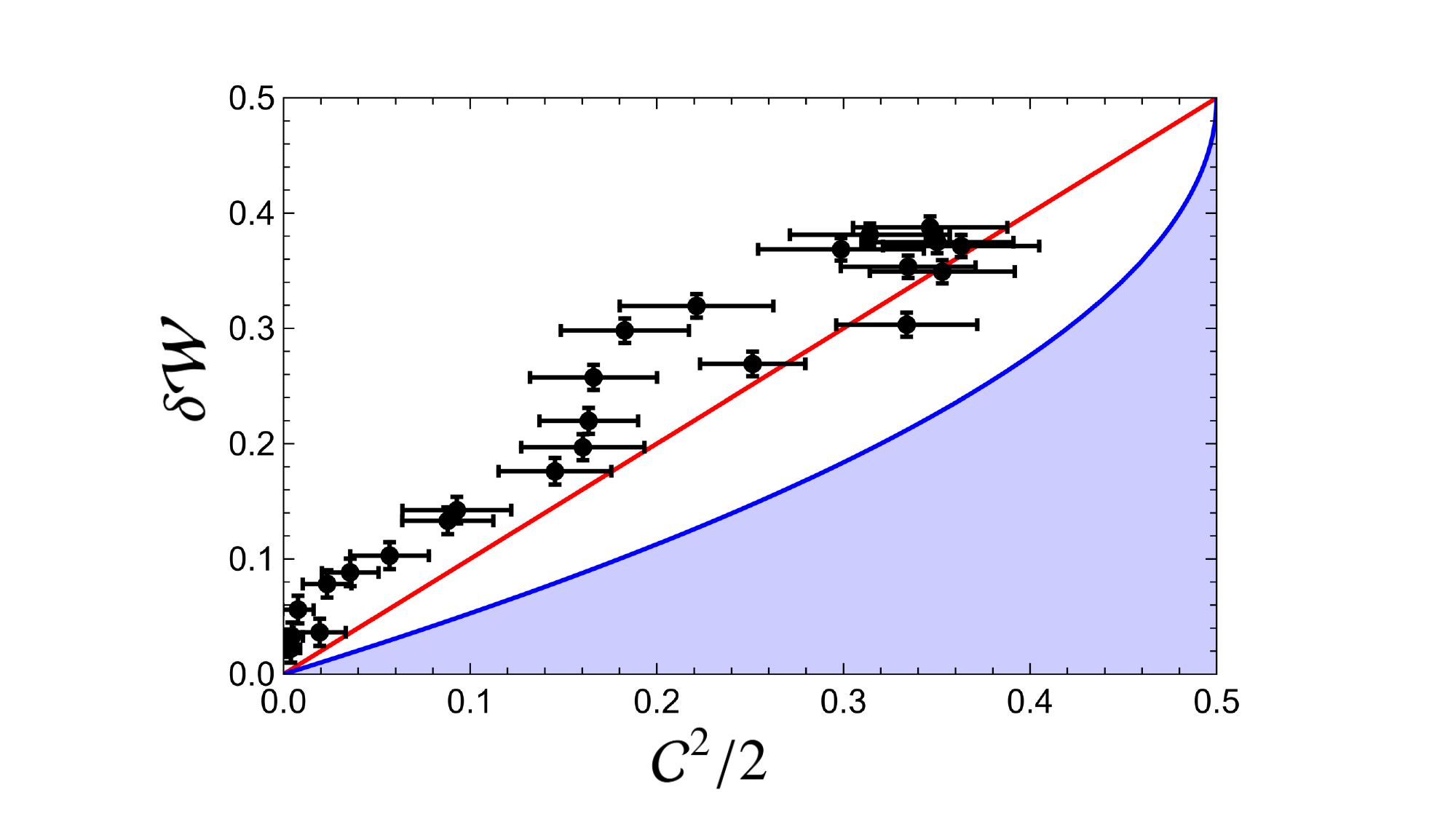}    \caption{Characterisation of the demonic energy gain $\delta\mathcal{W}$: The values for $\delta\mathcal{W}$ computed from the final populations shown in Fig. \ref{fig:prob_theta} and computed according to Eqs. \ref{eq:Wf},\ref{eq:Wi} are plotted versus the concurrence $\mathcal{C}(\theta)^2/2$ as shown in Fig. \ref{fig:fig3}.  We also indicate the theory prediction Eq. \ref{eq:dWvsC} (red) and the lower bound Eq. \ref{eq:dWlowerBound} (blue).}
    \label{fig:dw_c}
\end{figure}

A separate series of measurements is carried out to analyze the relevant properties of the resource state via quantum state tomography. The initial state Eq. \ref{eq:initialstate} is prepared by executing the circuit shown in Fig. \ref{fig:circuits} up to and including the parametric controlled-$Y$ gate for all values of $\theta$. For each setting, one of nine Pauli gates for measuring both qubits along the respective axes. For each measurement setting, 100 shots are acquired, the resulting population estimates are used for reconstructing the density matrix $\hat{\rho}$ of the resource state via linear inversion. This admits the computation of estimates for the concurrence $\mathcal{C}(\theta)$ via Eq. \ref{eq:concurrence} as well as the purity $\mathcal{P}(\theta)=\text{Tr}(\hat{\rho}^2)$. The resulting values for the concurrence and purity are shown in Fig. \ref{fig:fig3}. Confidence intervals for all quantities are computed via parametric bootstrapping: For each value of $\theta$, 500 artificial population data sets are computed from the reconstructed density matrix by sampling from the multinomial distribution determined by its diagonal. This data in turn is used to again compute density matrices via linear inversion, from which the metrics are computed and used to provide a measure of statistical spread. Note that density matrices obtained from linear inversion of finite measurement data are not necessarily positive, thus unphysical properties such as purity estimates $\mathcal{P}>1$ can be obtained. The results in Fig. \ref{fig:fig3} show that we can claim the resource states to be pure, justifying the finding that the demonic gain is determined by the concurrence. Fitting the concurrence values to $\mathcal{C}(\theta)=\mathcal{C}_0 \cos\theta$ reveals a consistent reduction factor $\mathcal{C}_0=0.87(2)$ which is due to gate errors (see below). \\

The measured demonic gain versus the estimated concurrence squared is shown in Fig. \ref{fig:dw_c}. Within statistical accuracy, the energy gain values match  the concurrence values according to Eq. \ref{eq:dWvsC} up to an offset, revealing a systematic deviation which can be explained by a microscopic gate error model. 
We empirically model imperfections of the controlled-$Z$ gates \cite{Hilder2022}: Taking into account that if such gates are carried out after shuttling operations, residual ion motion can lead to gate phases deviating from the preset value $\pi/2$. Simulating the entire circuit using qiskit \cite{QiskitCommunity2017}, we find the best match to the measured populations for gate phase deviations $\delta\Phi=\frac{\pi}{2}\times\{0.009(2), 0.068(9), 0.165(8)\}$, pertaining to the three controlled-$Z$ gates used in the circuit. As shown in Fig. \ref{fig:prob_theta} (solid lines), this provides a good agreement with the measurement populations. As the concurrence values fall short of the ideal values by the factor $\mathcal{C}_0$, the demonic gain values shown in Fig. \ref{fig:dw_c} systematically exceed the theory prediction Eq. \ref{eq:dWvsC}.\\

\textit{Discussion and outlook - } we have experimentally realized an agent-demon protocol, where the demon employs real-time feedback on a microscopic system, conditioned on measurement outcomes, to optimize extractable energy. Our results corroborate the finding that the energy gain that a party can obtain by applying an optimal feedback policy within a micropscopic system is governed by the entanglement between the subsystems. This work opens up perspectives for further studies of energy extraction in microscopic systems. In particular, it would be interesting to study non-isodimensional subsystems to explore the intermediate regime between a well-defined microscopic subsystem and a reservoir. Furthermore, it would be interesting to study how classical and non-classical correlations and feedback protocols can affect the extractable work from subsystems as determined by the ergotropy \cite{Allahverdyan2004,Francica2024}.\\

\textit{Acknowledgements - }
FSK and UGP acknowledge funding from DFG with research unit \textit{Thermal Machines in the Quantum World} (FOR 2724), from the EUH2020-FETFLAG-2018-03 under Grant Agreement no.820495 and by the Germany ministry of science and education (BMBF) within IQuAn. JG acknowledges support by the EPSRC-SFI joint project QuamNESS and by SFI under Frontier For the Future Program.  J.G. is supported by a SFI- Royal Society University Research Fellowship. MJK acknowledges the financial support from a Marie Sklodwoska-Curie Fellowship (Grant No. 101065974).

\bibliography{lit}

\end{document}